\documentstyle[prl,aps]{revtex}

\begin{document}

\twocolumn[\hsize\textwidth\columnwidth\hsize\csname
@twocolumnfalse\endcsname

\title{Effect of disorder on the vortex-lattice melting transition}
\author{E. A. Jagla and C. A. Balseiro}
\address{{\it Comisi\'on Nacional de Energ\'\i a At\'omica}\\
{\it Centro At\'omico Bariloche and Instituto Balseiro}\\
{\it 8400, S. C. de Bariloche, Argentina}}
\maketitle

\begin{abstract}
We use a three dimensional stacked triangular network of Josephson 
junctions 
as a model for the study of vortex structure in the mixed state of 
high $T_c$ superconductors. 
We show that the addition of disorder destroys the first order melting
transition occurring for clean samples. The melting transition  splits in two different 
(continuous) transitions, ocurring at temperatures 
$T_i$ and $T_p$ ($>T_i$). At $T_i$ the
perpendicular-to-field superconductivity is lost, and at $T_p$ the parallel-to-field
superconductivity is lost. These results agree well with recent experiments 
in $YBaCuO$.
\end{abstract}


\vskip2pc] \narrowtext

Although the existence of a melting transition of the vortex structure of
high $T_c$ superconductors is now accepted, the nature of the transition and
its dependence on different parameters is not completely clear in spite of
the big deal of theoretical and experimental work performed \cite{blatter}.
Experimentally, it is known that some external parameters can have strong
effect on the behavior of the system. Some of these parameters are the
magnetic field, the anisotropy, and the quenched disorder. It was found that
in some cases (low fields and weak disorder) the transition is first order 
\cite{exper} whereas in other cases (high fields or strong disorder) it is
continuous \cite{exper2,exper3}.

Theoretical studies on the nature of the transition focus on the problem of
whether it is a single transition at which superconducting coherence is lost
along all directions of the sample, or two successive transitions where
perpendicular and parallel to field coherence is lost in two sequential
steps \cite{teoria}. In addition, different possibilities for the transition
such as first order, second order, or crossovers have been proposed.
Theoretical treatments have been based on models that can usually account
only for part of the experimental data. The results depend on the parameters
used and sometimes they look contradictory \cite{teitcar}.

In this paper we fill part of the gaps between theory and experiment by
studying the three dimensional (3D) Josephson junction array (JJA) model in
the presence of an external magnetic field and with different kinds of
disorder. We show that when temperature increases this model gives different
results depending on the disorder: for low disorder we obtain a single {\em %
first order} transition where superconducting coherence is lost in all
directions, whereas for high disorder superconducting coherence is lost in
two successive {\em continuous} steps, first perpendicularly and then
parallel to the applied magnetic field. This behavior is closely related to
that found in experiments \cite{expern}.

The equations for the 3D JJA model are\cite{jb1} 
\begin{equation}
\label{eq1}j^{ii^{\prime }}=I_c^{ii^{\prime }}\sin \left( \varphi ^i-\varphi
^{i^{\prime }}-A^{ii^{\prime }}\right) +\frac{\phi _0}{2\pi R_0}\frac{%
\partial (\varphi ^i-\varphi ^{i^{\prime }})}{\partial t}+\vartheta
^{ii^{\prime }}(t)\text{ } 
\end{equation}
\begin{equation}
\label{eq2}\sum_{\{i^{\prime }\}}j^{ii^{\prime }}=j_{ext}^i, 
\end{equation}
where $\varphi ^i(t)$ is the phase of the superconducting order parameter, $%
I_c$ and $R_0$ are the critical current and normal resistance of each
junction, and $\phi _0$ is the flux quanta. Eq. (\ref{eq1}) gives the
current $j^{ii^{\prime }}$ between nearest neighbors nodes $i$ and $%
i^{\prime }$. Here $A^{ii^{\prime }}$ is the vector potential of the
external magnetic field, and $\vartheta ^{ii^{\prime }}(t)$ is an
uncorrelated gaussian noise which incorporates the effect of temperature.
Eq. (\ref{eq2}) ensures the current conservation on each node, and $%
j_{ext}^i $ is the external current applied at node $i$. We study this model
on a stack of triangular two-dimensional planes coupled by vertical links.

In all the simulations shown below the external current $j_{ext}$ (which is
used to calculate resistivities) is about $\sim 10^{-2}$ of the Josephson
junction critical current in the corresponding direction. If we want to
minimize the pinning effect of the subjacent JJA lattice we have to use a
magnetic field $H$ as low as possible. However, a low value of $H$ increases
the sample size needed to get good statistics. We have used a magnetic field 
$H$ equal to $1/6$ flux quanta per plaquette. This value of field generates
an Abrikosov lattice which is commensurate with the subjacent triangular JJA
lattice and is a compromise between low fields and not too long computation
time\cite{huse}. Self-inductance effects are not considered -this is
equivalent to take the magnetic penetration depth $\lambda \rightarrow
\infty $. When we simulate anisotropic systems we diminish the {\it c} axis
mean critical current of the junctions and at the same time increase the 
{\it c} axis elemental resistance by the same factor $\eta ^2$ ( $\eta
^2=\left\langle I_c^{\parallel }\right\rangle /\left\langle I_c^{\perp
}\right\rangle $ where $\parallel $ and $\perp $ indicate parallel an
perpendicular to the {\it ab }plane, and $\left\langle ...\right\rangle $
indicates the mean value on the sample). Boundary conditions (BC) are used
as follows: when external current is applied in the $x$ direction we take
open BC for the $x$ direction and periodic BC for $y$ and $z$ directions;
when external current is applied in the $z$ direction we take pseudo
periodic BC for the $z$ direction \cite{jb2} and periodic BC for $x$ and $y$
direction. In all cases the temperature is measured in units of the mean
Josephson energy of the in-plane junctions.

\noindent 
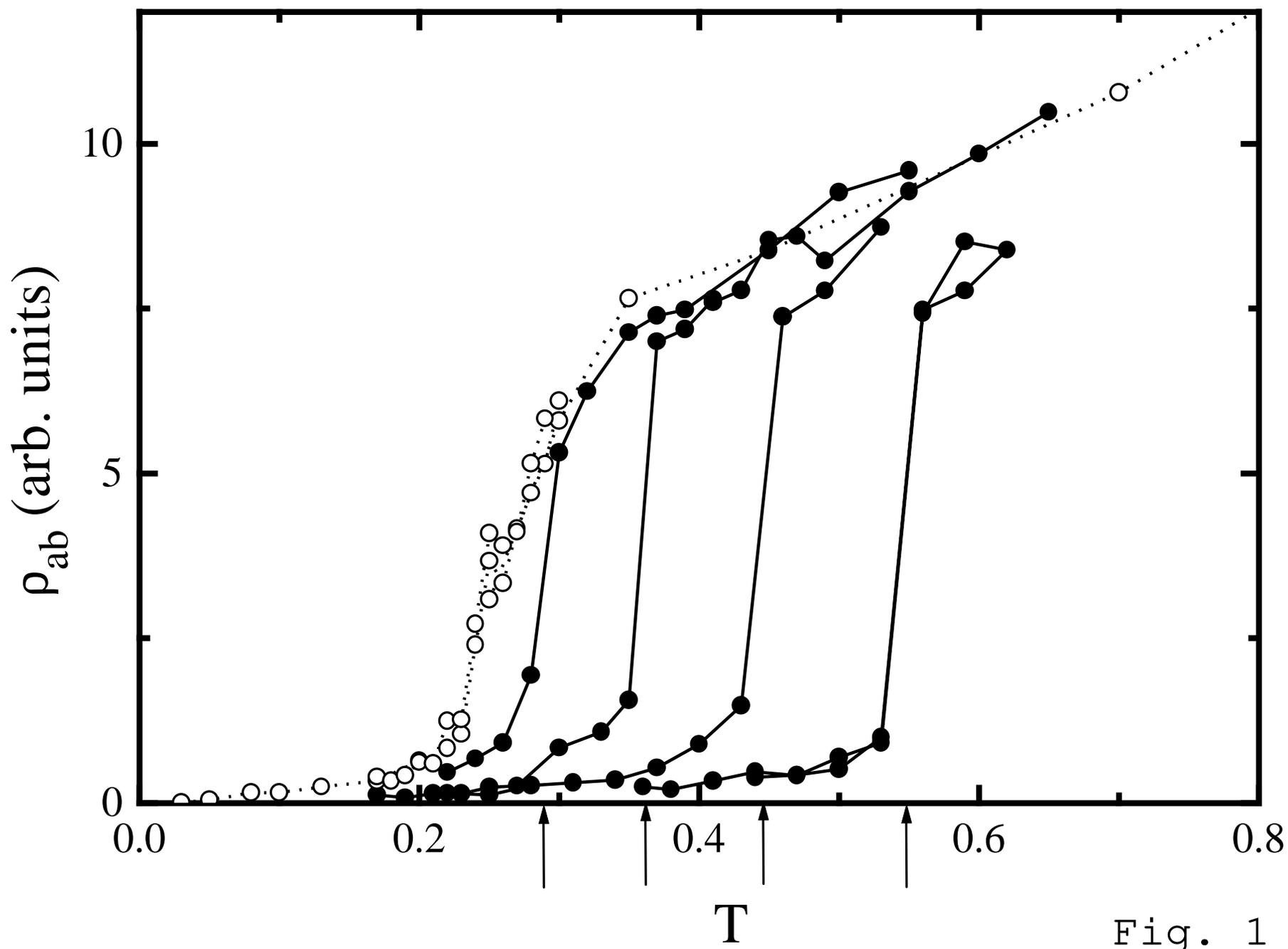
\begin{figure}
\hspace{-0.4cm}
\begin{picture}(8,6)
\end{picture}
\vspace{6cm}
\caption[fig1]
{Resistivity along the {\it ab} plane for a decoupled system of 
$18 \times 18 \times 18$ planes (open symbols), and for a three dimensional 
system with different anisotropies (full symbols, anisotropies $\eta ^2=$ 5, 10, 20, 
and 50 from right to left). The first order melting temperatures $T_m$ for the different
anisotropies  are indicated by arrows. Temperature is measured in units of the Josephson
energy of the in-plane junctions.}
\label{anis}
\end{figure}

We consider first the behavior of a system of uncoupled (triangular) planes
when the {\it c} coupling is turned on. It has been found that a single
plane has a weakly first order transition\cite{yu}. Our results for the
resistivity of this system are shown in Fig. \ref{anis} (open symbols).
Within the numerical precision we cannot distinguish from the behavior of
the resistivity between a continuous transition and a weak first order
transition occurring at $T\sim 0.23$ \cite{nota}. We will refer to this type
of behavior as a {\it continuous} transition, indicating that no
discontinuities are observed in the $\rho (T)$ curve. Both continuous
(second order) phase transitions and crossovers are in this category. When
the coupling is increased we clearly see a well defined transition
temperature where the resistivity has a jump. This is a {\em first order }%
transition as found in \cite{huse} and \cite{mingo}. In addition, the {\it c}
axis resistivity (not shown) has also a jump at the transition temperature,
indicating that the superconductor coherence is lost {\em discontinuously }%
and at the same temperature $T_m$ in all directions. Hysteresis loops, which
for the sake of clarity are not shown in the figure, are observed for all
the cases studied but for the uncoupled planes. We conclude that the {\it c}
axis coupling transforms a continuous transition into a first order
transition \cite{nota2}. Note that for high anisotropies, the resistivity
for $T>T_m$ is very close to the value corresponding to decoupled planes,
whereas for $T<T_m$ it is clearly different from that value. This indicates
that an {\em effective decoupling} of the planes is occurring at $T_m$. As
we argue below, the nature of the first order transition in thick samples is
different from that of an isolated plane.

It has been found experimentally that disorder alters the previous picture.
In fact, for samples with twin boundaries the transition in $\rho _{ab}$ is
continuous\cite{expern}. To study the effect of impurities in our model we
allow the critical currents $I_c$ of each junction to vary randomly between
two fixed values $I_c^{\min }$ and $I_c^{\max }$. This generate pinning
centers close to the junctions with low critical currents. All the results
shown in the simulations with disorder correspond to $D\equiv \left(
I_c^{\max }-I_c^{\min }\right) /\left( I_c^{\max }+I_c^{\min }\right) =0.5$.

\noindent 
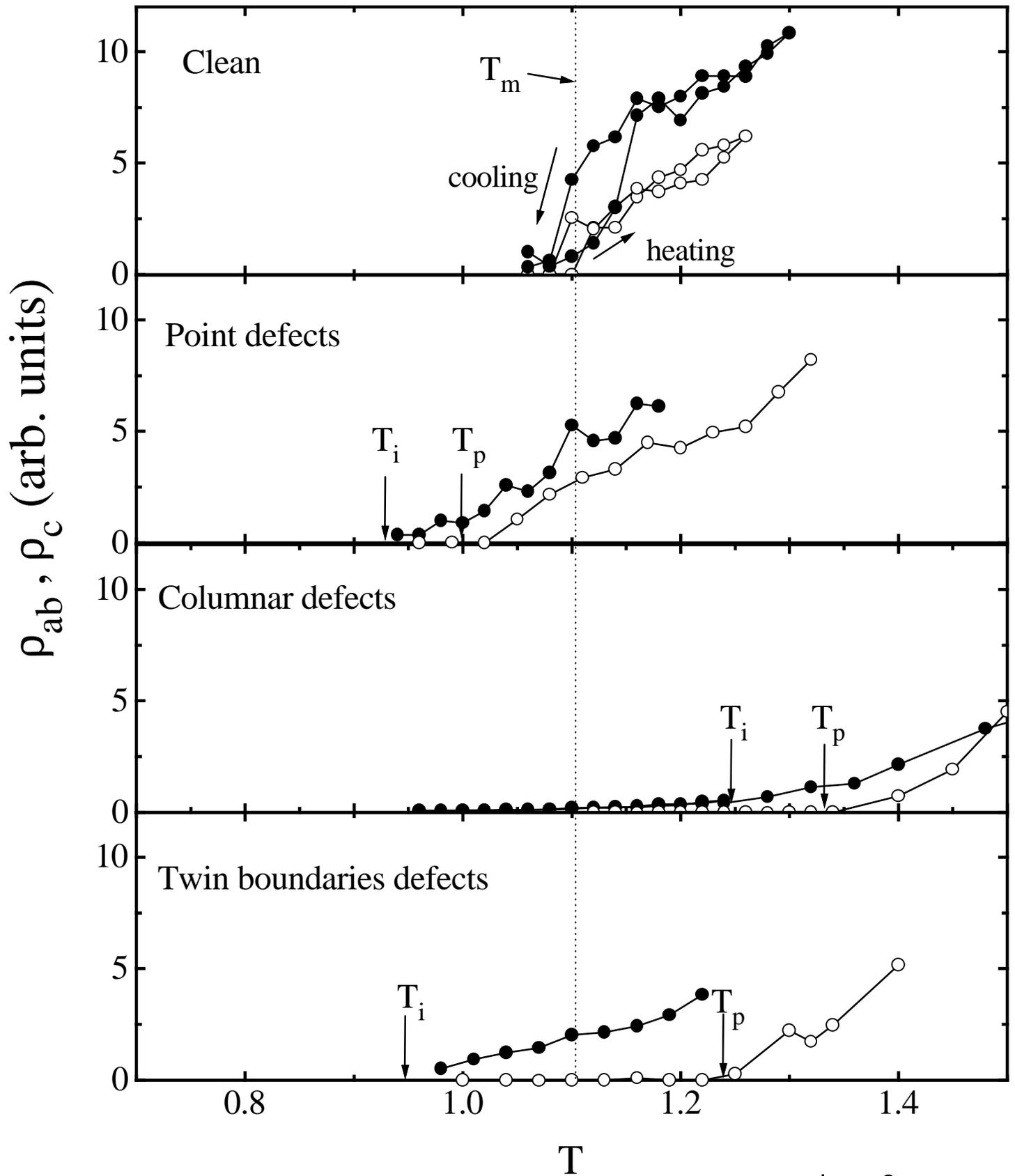
\begin{figure}
\hspace{-0.4cm}
\begin{picture}(8,12)
\end{picture}
\vspace{12cm}
\caption[fig2]
{Resistivities along the {\it ab} plane (full symbols) and along the {\it c} direction
for an isotropic $18 \times 18 \times 18$ sample and different configurations of
the disorder. For the clean sample hysteresis loops around $T_m$ upon heating and cooling 
are shown. For the disordered case the approximate values of $T_i$ and $T_p$ are
indicated.}
\label{roab}
\end{figure}

The effect of disorder depends on its spatial correlation. We consider three
different possibilities: point (uncorrelated), columnar (correlated along 
{\it c} axis), and planar (correlated in `twin boundaries' perpendicular to
the applied current) defects. For the uncorrelated disorder no correlation
on the values of $I_c$ for different junctions are set in. For the columnar
pinning all junction having the same {\it ab} plane coordinates have the
same critical currents, whereas there is no correlation for junctions with
different {\it ab} plane coordinates. For the twin boundaries case al
junctions with the same $x$ coordinate ($j_{ext}\parallel x$) have the same
critical currents, whereas there is no correlation for junctions with
different $x$ coordinates.

The results for $\rho _{ab}$ obtained for the different types of disorder
are illustrated in Fig. \ref{roab} (full symbols) for an isotropic ($%
\left\langle I_c^{\parallel }\right\rangle =\left\langle I_c^{\perp
}\right\rangle $) sample. For the clean sample the hysteresis loop is shown,
indicating again a first order transition at $T=T_m$. Although hysteresis in
the resistivity is not an unequivocal indication of a first order
transition, we checked that the transition is in fact first order by using
the histogram technique\cite{leek,huse}. In addition other indication of
loose of coherence (in particular the helicity modulus) give results which
are in agreement with the conclusions obtained from resistivity
measurements. In all disordered cases the transition in $\rho _{ab}$ is
continuous, and we denote the transition temperature as $T_i$. Point defects
and twin boundaries decrease the transition temperature, whereas columnar
defects increase it. This is consistent with the fact that columnar pinning
generates the strongest pinning.

It is instructive to study the behavior of $\rho _c$ with disorder. In Fig. 
\ref{roab} (open symbols) we show the values of $\rho _c$ for the same
disorder configurations as before. The main result is that $\rho _c$ also
becomes continuous in the presence of disorder, and the transition
temperature (that we denote $T_p$\cite{jb1,jb2} and experimentalists $T_{th}$
\cite{exp1}) is higher than the corresponding $T_i$. The temperature $T_p$
is higher than $T_m$ for samples with columnar defects and twin boundaries,
whereas is lower than $T_m$ for point defects. This indicates that point
defects diminish the coherence along the {\it c} direction, whereas the
other two types of defects enhance it. Note also that the absolute
difference $T_p-T_i$ is the highest for the twinned sample. This is due to
the fact that planar defects parallel to the vortex movement have opposite
effects on $T_i$ and $T_p$: $T_i$ diminishes because there are paths of easy
movement for the vortices, and the rigidity of the ideal Abrikosov lattice
has been destroyed, and $T_p$ increases because the defects are {\it c} axis
correlated, and thus they enhance the coherence along the {\it c} direction.

We found that if the value of the disorder $D$ is smaller than a critical
value $D_c$ ($D_c\sim 0.3$ for isotropic samples) the first order transition
persist and the coherence in all directions is lost at the melting
temperature. Strong disorder ($D>D_c$) is necessary to produce the behavior
of Fig. \ref{roab}.

Our results are in complete agreement with recent results on {\it YBaCuO}
samples, where a first order transition is observed in clean samples and two
transitions are observed with $T_i<T_m<T_p$ in twinned samples \cite{expern}%
. It has also been shown that point defects destroy the first order
transition \cite{exper2}, but unfortunately pseudo DC transformer or direct 
{\it c} axis resistivity measurements were not performed on these samples.

We checked that the transitions at $T_i$ and $T_p$ occurring in our
triangular samples with disorder have the same features described in
references \cite{jb1,jb2,jb3} for the case of a square lattice of junctions:
the transition at $T_i$ is a thermal depinning of vortex lines, which is
probably related to the large amount of disorder \cite{rao}; the transition
at $T_p$ is a percolation phase transition of the vortex lattice
perpendicularly to the applied field. Also we have checked that around $T_p$%
, $\rho _c(T)$ satisfies the percolation scaling laws of \cite{jb3}. In
addition, the structure of the lattice for $T_i<T<T_p$ is that of a
disentangled vortex liquid. The range $T_p-T_i$ in which this disentangled
liquid exists shrinks to zero when increasing the thickness of the sample%
\cite{jb2,jb3}.

When two transitions at different temperatures $T_i$ and $T_p$ occur, it is
natural that both of them are continuous: at $T_i$ the system changes from a
solid to a liquid of {\it c} axis correlated (and thus effectively two
dimensional) vortices. This transition is qualitatively similar to the
fusion of a two dimensional system, and it is expected to be continuous. At $%
T_p$ the transition is driven by a percolation of vortex loops, which is a
second order (continuous) phase transition.

Experimental and theoretical evidence suggest that the transition is first
order {\em only} when the superconducting coherence is lost in all
directions at the same temperature. This suggests that the first order
melting in clean samples is a consequence of the interplay between two
different transitions: the depinning of single vortex lines -which drives
the crossover occurring at $T_i$-, and the percolation of vortex loops
between planes -which is responsible for the appearance of dissipation along
the {\it c} direction at $T_p$. These two transitions may cooperate and
merge onto a single one for clean samples: when percolation occurs the
planes decouple, and this reduces the effective value of $T_i$. In turn,
when vortices depin, vortex loops between planes are more easily generated
due to the screening effect of mobile vortices, thus reducing $T_p$. The
combination of these two effects can generate an instability that drives the
transition first order \cite{jbtbp}.

As we said above, the merging of $T_i$ and $T_p$ onto a single first order
melting temperature $T_m$ is observed only for clean samples. However, if
the thickness of the system is lower than a minimum value two different
transitions at $T_i$ and $T_p$ are observed. In particular, in our
simulations the thickness of the system has to be greater than ten planes on
isotropic samples in order to obtain a first order transition. For thinner
samples we observe two separate and continuous transitions. One can make a
rough estimate of the dependence of the critical thickness on field in the
following way. Previous results \cite{jb2,jb3} indicate that $T_p$ depends
on thickness as 
\begin{equation}
\label{qq}k_BT_p\sim \Delta /\ln \left( L_c/d\right) , 
\end{equation}
with $d$ a distance of the order of the interlayer spacing, $L_c$ the
thickness of the sample, and $\Delta $ an energy which is the order of the
energy necessary to make two nearest vortices touch each other. Being $%
\Delta $ proportional to the distance between vortices we can write $%
\Delta=\epsilon _0\phi _0\sqrt{H}$, with $\epsilon _0$ an energy scale
factor related to the linear energy density of a vortex line. To be able to
merge onto a single transition, $T_i$ and $T_p$ should be of the same order.
Having into account that $T_i$ is rather thickness independent\cite{jb2,jb3}
we obtain the minimum value $L_{\min }$ necessary to have a first order
transition as 
\begin{equation}
L_{\min }=d\exp \left( \epsilon _0\phi _0/\sqrt{H}k_BT_i\right) . 
\end{equation}

This expression shows that $L_{\min }\rightarrow \infty $ for $H\rightarrow
0 $, indicating that very thick samples are necessary to observe the first
order transition at very low fields. This is probably not an experimental
limitation due to the small value of $d$, but it should be taken into
account in numerical simulations at low fields.

We have concentrated on the effect of disorder on the melting transition at
a fixed value of the external field. Finding the behavior of the system when
changing the magnetic field $H$ is difficult because of commensurability
effects between the vortex lattice and the mesh. However, we can choose to
change at the same time the magnetic field and the discretization parameter,
in such a way that for any real field we have 1/6 flux quanta per plaquette.
The change in the discretization produces a change in the anisotropy $\eta $
and the disorder of the system. For a real change in $H$ of the form $%
H\rightarrow H\delta $ with $\delta \simeq 1$, the anisotropy changes as $%
\eta \rightarrow \eta \delta ^{1/2}$ \cite{chentei}. The change in the
disorder may depend on the particular realization. In the simplest case of
uncorrelated disorder it can be shown that $D$ changes as $D\rightarrow
D\delta ^{1/2}$. We conclude that we can study the $H-T$ phase diagram of
high-$T_c$'s by analyzing the $\eta -D-T$ phase diagram along lines of fixed
value of $\eta /D$. Preliminary results indicate that the overall properties
of the $H-T$ phase diagrams of {\it YBaCuO }and {\it BiSrCaCuO} are well
reproduced \cite{jbtbp}. In particular, from these arguments and Fig. \ref
{anis} we can conclude that the jump in the resistivity at the first order
melting should decrease when increasing $H$.

In summary, we have presented results obtained using the three dimensional
JJA model that reproduce very well the behavior of the vortex lattice in
high-$T_c$ superconductors. For clean samples the system has a first order
melting transition at temperature $T_m$, at which the superconducting
coherence is lost discontinuously in all directions. In the presence of
disorder the transition separates in two continuous ones occurring at
temperatures $T_i$ and $T_p$ ($>T_i$). At $T_i$ perpendicular-to-field
correlation is lost due to depinning of vortices from the pinning centers.
At $T_p$ parallel-to-field correlation is lost due to a percolation
transition of the vortex lattice. In addition, relative values of $T_i$, $T_p
$, and $T_m$ depend on the kind of disorder considered.

We acknowledge useful discussions with D. L\'opez, E. Righi, and F. de la
Cruz. We also thank D. Dom\'?nguez for critical reading of the manuscript.
E.A.J. is supported by CONICET, and C.A.B. is partially supported by CONICET.


\begin{references}
\bibitem{nota}In all cases a low temperature tail is observed in 
$\rho _{ab}(T)$ which is due to surface effects.
\bibitem{nota2} Note that the latent heat per vortex per plane reported in \cite{huse} 
is one order of magnitude greater than the corresponding to a single plane \cite{yu}.
\end{references}
\end{document}